\documentclass[letterpaper,twocolumn,10pt]{article}
\usepackage{uformat}
\usepackage[titletoc]{appendix}
\usepackage{amsmath}
\usepackage{algorithm}
\usepackage{algpseudocode}
\usepackage{url}
\usepackage{graphicx}
\usepackage{courier}
\usepackage{xcolor}
\usepackage{color}
\usepackage{listings}
\usepackage{caption}
\usepackage{subcaption}
\usepackage{multirow}
\usepackage{booktabs}
\usepackage{dcolumn}
\usepackage{outlines}
\usepackage{amssymb}
\newcolumntype{d}[1]{D{.}{.}{#1}}
\usepackage{mwe}
\usepackage{adjustbox} 
\usepackage{tikz}
\usepackage{pgfplots}
\pgfplotsset{width=6cm,compat=1.12}
\usepgfplotslibrary{fillbetween}

\usepackage{array}
\newcolumntype{C}[1]{>{\arraybackslash}p{#1}}

\newcommand{\method}{{\scshape Soteria}}

\title{\method{}: In Search of Efficient Neural Networks for Private Inference}

\author{Anshul Aggarwal, Trevor E. Carlson, Reza Shokri, Shruti Tople\\ National University of Singapore (NUS), Microsoft Research \\ \texttt{\{anshul, trevor, reza\}@comp.nus.edu.sg, shruti.tople@microsoft.com }}

\begin{document}

\maketitle

\begin{abstract}	
	ML-as-a-service  is gaining popularity where a cloud server hosts a trained model and offers prediction (inference) service to users. In this setting, our objective is to protect the confidentiality of both the users' input queries as well as the model parameters at the server, with modest computation and communication overhead.  Prior solutions primarily propose fine-tuning cryptographic methods to make them efficient for known {\em fixed} model architectures.  The drawback with this line of approach is that the model itself is never {\em designed} to operate with existing efficient cryptographic computations.  We observe that the network architecture, internal functions, and parameters of a model, which are all chosen during training, significantly influence the computation and communication overhead of a cryptographic method, during inference.  Based on this observation, we propose \method{} --- a training method to construct model architectures that are by-design efficient for private inference.  We use neural {\em architecture search} algorithms with the dual objective of optimizing the accuracy of the model and the overhead of using cryptographic primitives for secure inference.  Given the flexibility of modifying a model during training, we  find accurate models that are also efficient for private computation.  We select garbled circuits as our underlying cryptographic primitive, due to their expressiveness and efficiency, but this approach can be extended to hybrid multi-party computation settings.  We empirically evaluate \method{} on MNIST and CIFAR10 datasets, to compare with the prior work. Our results confirm that  \method{} is indeed effective in balancing performance and accuracy.  
\end{abstract}

\section{Introduction}
\label{sec:intro}

Machine learning models are susceptible to several security and privacy attacks throughout their training and inference pipelines. Defending each of these threats  require different types of security mechanisms.  The most important requirement is that the sensitive input data as well as the trained model parameters  remains confidential at all times.  In this paper, we focus on private computation of inference over deep neural networks, which is the setting of machine learning-as-a-service.  Consider a server that provides a machine learning service (e.g., classification), and a client who needs to use the service for an inference on her data record.  The server is not willing to share the proprietary machine learning model, underpinning her service, with any client.  The clients are also unwilling to share their sensitive private data with the server.  We consider an honest-but-curious threat model.  In addition, we assume the two parties do not trust, nor include, any third entity in the protocol.  In this setting, our first objective is to design a secure protocol that protects the confidentiality of client data as well as the prediction results against the server who runs the computation.  The second objective is to preserve the confidentiality of the model parameters with respect to the client.  We emphasize that the protection against the indirect inference attacks that aim at reconstructing model parameters~\cite{tramer2016stealing} or its training data~\cite{shokri2017membership}, by exploiting model predictions, is not our goal.

A number of techniques provide data confidentiality while computing and thereby allow {\em private computation}.  The techniques include computation on trusted processors such as Intel SGX~\cite{hunt2018chiron, ohrimenko2016oblivious}, and computation on encrypted data, using homomorphic encryption, garbled circuits, secret sharing, and hybrid cryptographic approaches that jointly optimize the efficiency of private inference on neural networks~\cite{YaoGC,gentry,yao1982protocols,brakerski2011fully,beaver1991efficient}.  To provide private inference with minimal performance overhead and accuracy loss, the dominant line of research involves  adapting cryptographic functions to (an approximation of) a given {\em fixed} model~\cite{MiniONN, SecureML, Delphi, DeepSecure, XONN, EzPC, Gazelle, Chameleon}. 
However, the alternative approach of {\em searching or designing} a network architecture for a given set of efficient and known cryptographic primitives is unexplored in the literature.

{\bf Our Contributions.} In this work, we approach the problem of privacy-preserving inference from a novel perspective. Instead of modifying cryptographic schemes to support neural network computations, we advocate modification of the training algorithms for efficient cryptographic primitives. Research has shown that  training algorithms for deep learning are inherently flexible with respect to their neural network architecture. This means that different network configurations can achieve similar level of prediction accuracy. We exploit this fact about deep learning algorithms and investigate the problem of optimizing deep learning algorithms to ensure efficient private computation. 

To this end, we present \method{} ---  an approach for constructing deep neural networks optimized for performance, accuracy and confidentiality. Among all the available cryptographic primitives, we use garbled circuits as our main building block to address the {\em confidentiality} concern in the design of \method{}. Garbled circuits are known to be efficient and allow generation of constant depth circuits even for non-linear function which is not possible with other primitives such as GMW or FHE.  We show that neural network algorithms can be optimized to efficiently execute garbled circuits while acheiving high accuracy guarantees. We observe that the efficiency of evaluating an inference circuit depends on two key factors: the model parameters and the network structure.  With this observation, we design a regularized architecture search algorithm to construct neural networks. \method{} selects optimal parameter sparsity and network structure with the objective to guarantee acceptable {\em performance} on garbled circuits and high model {\em accuracy}.

\section{Selecting the Cryptographic Primitive} \label{sec:selection}

\begin{table}[t!]
		\centering
	\begin{tabular}{@{}|l|c c c c c|@{}}
		\toprule
		& PHE & FHE & SS & GMW & GC \\ \midrule
		Expressiveness                                                            & $\times$   & $\checkmark$   & $\checkmark$  & $\checkmark$   & $\checkmark$  \\ \midrule
		Efficiency                                                                &  $\checkmark$   & $\times$   & $\checkmark$  & $\checkmark$   & $\checkmark$  \\ \midrule
		\begin{tabular}[c]{@{}l@{}}Communication \\ (One time setup) \end{tabular} & $\checkmark$   & $\checkmark$   & $\times$  & $\times$   & $\checkmark$  \\ \bottomrule
	\end{tabular}
	\caption{Properties of secure computation cryptographic primitives: Partially and fully homomorphic encryption schemes (PHE, FHE), Goldreich-Micali-Widgerson protocol (GMW), arithmetic secret sharing (SS), and Yao's garbled circuit (GC).}
	\label{table:cryptoprimitives}
\end{table}

In designing \method{}, we make several design choices with the goal of achieving efficiency. The most important among them is the selection of the underlying cryptographic primitive to ensure privacy of data.  Several cryptographic primitives such as partially homomorphic encryption schemes (PHE) and fully homomorphic encryption schemes (FHE), Goldreich-Micali-Widgerson protocol (GMW), arithmetic secret sharing (SS), and Yao's garbled circuit (GC) have been proposed to enable two-party secure computation.  Each of these primitives perform differently with respect to the factors such as efficiency, functionality, required resources and so on. PHE schemes allow either addition or multiplication operations but not both on encrypted data~\cite{paillier1999public, elgamal1985public}. In contrast, FHE schemes enable both addition and multiplication on encrypted data~\cite{gentry,FHE, van2010fully, brakerski2011fully} but incur huge performance overhead. SS involves distributing the  secret shares among non-trusting parties such that any operation can be computed on encrypted data without revealing the individual inputs of each party~\cite{beaver1991efficient}.  GMW~\cite{GMW} and GC~\cite{yao1982protocols} allow designing boolean circuits and evaluating them between a client and a server. The differences between these schemes might make it difficult to decide which primitive is the best fit for designing a privacy-preserving system for a particular application.  Therefore, we first outline the desirable properties specifically for private neural network inference and then compare these primitives with respect to these properties (see Table~\ref{table:cryptoprimitives}). We select a cryptographic scheme that satisfies all our requirements.

{\bf Expressiveness.} This property ensures that the cryptographic primitive supports encrypted computation for a variety of operations. With the goal to enable private computation for neural networks, we examine the type of computations required in deep learning algorithms. Neural network algorithms are composed of linear and non-linear operations. Linear operations include computation required in the execution of fully-connected and convolution layers. Non-linear operations include activation functions such as Tanh, Sigmoid and ReLU. The research in deep learning is at its peak with a plethora of new models being proposed by the community to improve the accuracy of various tasks. Hence, we desire that the underlying primitive should be expressive with respect to any new operations used in the future as well.  PHE schemes offer limited operations (either addition or multiplication) on encrypted data. This limits their usage in applications that demand expressive functionalities such as neural network algorithms. Alternative approaches such as FHE, SS, GMW and GC protocols allow arbitrary operations.

{\bf Computation Efficiency.}  Efficiency is one of the key factors while designing a client-server application such as a neural network inference service on the cloud.  FHE techniques have shown to incur orders of magnitude overhead for computation of higher-degree polynmials or non-linear functions. Existing approaches using FHE schemes have restricted its use to compute only linear functions. However, most of the neural network architectures such as CNNs have each linear layer followed by a non-linear layer. To handle non-linear operations, previous solutions either approximate them to linear functions or switch to cryptographic primitives that support non-linearlity~\cite{Cryptonets, SecureML, Gazelle, Delphi}.  Approximation of non-linear functions such as ReLU highly impacts the accuracy of the model. Switching between schemes introduces additional computation cost which is directly proportional to the network size.  In comparison to FHE, research has shown that SS, GMW and GC schemes provide constructions with reasonable computation overhead for both linear and non-linear operations.

{\bf Communication Overhead.} The communication costs incurred for private computation contribute to the decision of selecting our cryptographic primitive, as the network should not become a bottleneck in the execution of the private machine learning as a service.  We expect the client and server to interact only once during the setup phase and at the end of the execution to receive the output. We aim to remain backward compatible to the existing cloud service setting where the client does not need to be online at all time between the request and response.  In contradiction to this property, the GMW scheme requires communication rounds proportional to the depth of the circuit. To evaluate every layer with an AND gate, the client and server have to exchange secrets among them forcing the client to be online throughout the execution. Similarly, construction of non-linear bitwise functions with arithmetic secret shares require communication rounds logarithmic to the number of bits in the input. This makes the use of these schemes almost infeasible in the cloud setting that have a high-latency network.  Unlike these primitives, Yao's garbled circuits combined with recent optimizations require an exchange of data only once at the beginning of the protocol. 

We select GC as our underlying cryptographic primitive in \method{} as it satisfies all the desired properties for a designing private inference for cloud service applications. 

\section{Garbled Circuit for Efficient Neural-Networks}\label{sec:gcbinary}

We investigate the problem of performing private inference on neural networks. Let $W$ be the model parameters stored on the server, $x$ be the client's input, $y$ be the expected output and $f$ is the inference function to be computed. Given this, we want to compute ${f(x;\theta)\rightarrow y}$. We aim for the following main goals:
\begin{itemize}
	\item {\em Confidentiality:}
	The solution should preserve confidentiality of the model parameters $\theta$ from the users and that of $x$ and $y$ from the server. We assume an honest-but-curious threat model. 
	\item {\em Accuracy:}
	The drop in accuracy of the privately computed inference function should be negligible as compared to the accuracy of the model on plaintext data.
	\item {\em Performance:}
	The private computation should demonstrate acceptable performance (runtime and communication) overhead.
\end{itemize}

\paragraph{\em Garbled circuits.} 
GC protocol allows construction of any function as a boolean circuit with a one time setup cost of data exchange~\cite{YaoGC}. In our setting, the client is the {\em garbler} and the server is the {\em evaluator}. In the setup phase, the client first transforms the function into a boolean circuit with two-input gates. The function (model architecture) and the circuit are known to both the parties, but its parameters and input are private.  The client then garbles the circuit.  This process involves creating a garbled computation table (GCT), which is an encrypted version of the truth table for the boolean circuit.  The entries for this table are randomly permuted, so that the order does not leak information. The client then shares the garbled circuit and its encrypted inputs to the circuit (binary values representing $x$) with the server. In the next phase, the parties perform an oblivious transfer (OT) protocol~\cite{ObliviousTransfer_paper}, so the server obtains the encryption of its inputs to the circuit (binary values representing $W$), without leaking information about its parameters to the client. Then, the server evaluates the circuit, and obtains the output value which is encrypted. The server transfers the output to the client which can match the encrypted values to their plaintext and obtain $f(x;\theta)$. 

\paragraph{\em Performance.} 
In GC, the communication and computation overhead is directly dependent on the number of AND, OR gates in the boolean circuit. Prior research has proposed several techniques that make it free for the GC to execute XOR, XNOR and NOT gates~\cite{kolesnikov2008improved}.  Given this prior work, the communication overhead of the GC protocol for a given circuit is proportional to its security parameter and the number of non-XOR gates in the circuit.  The total runtime for evaluating a circuit is the sum of the time required during the {\bf online} (evaluation) and {\bf offline} (garbling and oblivious transfer) computation.

\paragraph{\em Efficient neural networks.}  
In \method{}, we leverage the above-mentioned properties of GC to design an optimized neural network algorithm. Neural network algorithms are shown to be flexible with respect to their architectures i.e., multiple models with different configuration can achieve a similar level of accuracy.  We take advantage of this observation and propose designing neural network {\bf architectures} that help optimize the performance of executing inference with garbled circuits.  The number of gates in a circuit corresponding to a neural network depends on its activation functions and the size of its parameter vector. 

Neural networks have shown to exhibit relatively high accuracy for various tasks even with low precision parameters.  {\bf Binary neural networks}~\cite{BNN} are designed with the lowest possible size for each parameter, i.e., one bit to represent $\{-1, +1\}$ values.  Using BNNs naturally aligns with our selected cryptographic primitive because each wire in garbled circuits represents 1 bit value (representing $-1$ in the model with $0$ in the circuit).  Binarizing the model parameters further allows us to heavily use the free XOR, XNOR and NOT gates in garbled circuits, thus minimizing the computation and communication overhead of private inference.  This has recently been shown in the performance evaluation of garbled circuits on binary neural networks~\cite{XONN}.  

In neural networks, {\bf linear functions} such as those used in the convolutional or fully connected layers form an important part of the network. These functions involve dot product vector multiplications.  Instead of using multiplications, this can be computed very efficiently using XNOR-popcount: 
${\mathbf{x} \cdot \mathbf{w} = 2\times\operatorname{bitcount}(\operatorname{xnor}(\mathbf{x}, \mathbf{w})) - N}$, where $N = |\mathbf{x}|$.  In binary neural networks, the output of activation functions is also binary.  But, the output of XNOR-popcount is not a binary number, thus, according to BNN algorithms one would need to compare it with $0$; positive numbers will be converted to $1$ and negative numbers will be converted to $0$. 

We can compute some {\bf non-linear functions} such as maxpool very efficiently in BNNs.  Max-pooling is a simple operation which returns the maximum value from a vector, which in the case of neural networks is usually a one-dimensional representation of a 2D max-pooling window. In binary neural networks, maxpool need to simply return $1$ if there is a $1$ in the vector. This is achieved by a logical OR-operation over the elements of the vector. 

To achieve a learning capacity for binary neural networks similar to full-precision models, we would need to scale up the the number of model parameters.  We can increase the number of kernels in a convolution layer and the number of nodes in a fully connected layer, by a given {\bf scaling factor}. This technique has been used in the prior work~\cite{XONN}, and enables learning more accurate models, however at the cost of increasing the number of computations in the network.

\section{\method{}} \label{sec:method}

All the techniques which we discuss in Section~\ref{sec:gcbinary}, can help in  reducing the overhead of the garbled circuit protocol on a neural network.  However, besides the size of model parameters, which is reduced in binary neural networks, the {\em model size} and its {\em structure} also play significant roles in determining the number of non-XOR gates in the garbled circuit of neural networks.  For example, the configurations of the convolutional layers directly affects the overhead of garbled circuits on neural networks.  Besides, not all model parameters are of the same value for the machine learning task, and models with the same structure but with larger sparsity can result in similar accuracy, but significantly lower overhead for private computation. 

In this paper, we design \method{} to automatically learn the model architecture and its connections so as to optimize the cost of private inference in addition to optimizing accuracy.  This is a different approach than simply fine-tunning or compressing a model, as we aim at including the cost of private computation as part of the objective of {\em architecture learning} and {\em parameter learning} of the model.  To this end, we build \method{} on top of two well-established classes of machine learning algorithms to search for the models that balance accuracy and performance: neural architecture search algorithms, and ternary neural network algorithms.

\subsection{Neural architecture search for constructing efficient models for private inference}

Architecture search algorithms for neural networks are designed to replace the manual design of complex deep models.  The objective is to learn the model structure for which we hope to obtain a high accuracy when trained on the training set.  A number of such algorithms are designed recently.  NAS~\cite{NAS_survey} is one of the first neural architecture search algorithms.  It comprises of three components --- a \textit{search space} which is the domain of architectures over which the search will be executed, the \textit{search strategy}, which defines how the search space has to be explored, and a \textit{performance estimator}, to estimate the performance of a particular discovered architecture on unseen data. Multiple techniques have been proposed  to minimize the computation cost of the search process, by tweaking the search strategy and the performance estimator such as ENAS and DARTS~\cite{nas_rl, enas, DARTS}. DARTS is a differentiable neural architecture search algorithm, which is orders of magnitude faster than other search algorithms ~\cite{DARTS}.  DARTS automatically constructs the model architecture by stacking a number of {\em cells}.  Each cell is a directed acyclic graph, where each node is a neural {\em operation} (e.g., convolution with different dimensions, maxpool, identity).  The architecture search algorithm learns the optimal construction of cells that would maximize the accuracy of the model on some validation set.  During the search algorithm, we use stochastic gradient descent to continuously update the probability of using different candidate operations for each connection in the internal graph of a cell.  These probabilities reflect the usefulness of {\em each} operation for different positions in the cell.  Let $\alpha_{o}^{(i,j)}$ be the fitting score associated with operation $o$ to connect nodes $i$ and $j$ in the directed acyclic graph inside the cell.  The probability of choosing a particular operation to connect node $i$ to $j$ is computed as a softmax of the score $\alpha_o^{(i,j)}$ over all possible operations.

In \method{}, we modify the computation of the $\alpha_o^{(i,j)}$ scores over candidate operations.  In order to include the cost of private inference, we penalize each operation proportional to its computation and communication overhead.  Let $\gamma(o)$ be the penalty or the cost function for an operation $o$. The penalty factor could be the normalized runtime and communication cost of an operation, which can be computed empirically on garbled operations.  In our experiments, we compute the penalty factor for different operations in Table~\ref{table:operationcost}.  We update the fitting score $\alpha_o^{(i,j)}$ by replacing it with $\alpha_o^{(i,j)} (1-\lambda \gamma(o))$, where $\lambda$ is our regularization term. Larger values of $\lambda$ would result in models that prefer training efficient models over accurate models.  

By regularizing the architecture search algorithm, we effectively guide the algorithm to identify a configuration for cells which optimize both model accuracy and performance of private inference.  This enables fine-tuning the model {\em before} being trained to be efficient on our cryptographic primitives.  As we balance the trade-off between accuracy and performance, \method{} can construct models which {\em by design} satisfy the requirements of our system.  In our experiments, we evaluate the performance of models under different values of $\lambda$, and how this factor can be used to balance different costs of confidentiality for neural networks. 

\subsection{Ternary (Sparse Binary) Neural Network} 

\label{section:tnn}
\label{sec:ternarization-of-weights}

For building a system that enables efficient private inference, we prefer to reduce the number of parameters in the network. One approach is to train a model and then compress the model, however, that might not result in the best construction of the model as far as the model accuracy is concerned. Besides, to be aligned with our approach of {\em constructing} model architectures, we would prefer to learn model structures which are sparse.  One well-established machine learning technique is to learn a model with ternary values ($-1, 0, +1$).  This effectively means that some of the network connections (parameters) are removed (for parameters with value $0$).  Ternary neural networks try to minimize the distance between the full precision model parameters and their ternary values~\cite{twn}.  

In building models for \method{}, we train models with ternary parameters and binary activation functions. This would enable us to still use the techniques for binary neural networks, as discussed in Section~\ref{sec:gcbinary}, however on a smaller circuit (due to the model's sparsity).  We incorporate ternary neural networks into our regularized architecture search algorithm to find cells containing only ternary convolution and max-pooling layers that operate on binary inputs and ternary parameters.

The procedure for converting full precision model parameters, during  training, to ternary involves comparing the model parameters with a threshold in the forward pass of the gradient descent algorithm. We follow the established algorithms in this domain~\cite{twn}.  For the parameters $W_i$ in an operation, we convert the parameter to $+1$ when it is larger than threshold $\Delta$, we set it to $-1$ if it is smaller than $-\Delta$, and we set it to $0$ otherwise. The threshold is computed as ${\Delta = \frac{0.7}{n} \sum_{i=1}^{n}\left|\mathrm{W}_{i}\right|}$.  Output of the functions are also binarized similarly, by comparing them with $0$, where positive values are converted to $+1$, and the negative values are converted to~$-1$.  All these transformations happen {\em during} the model training, so the final model is optimal given the ternary restrictions.  Besides, the training algorithm finds the optimal level of sparsity for the model which does not conflict with its accuracy.

\begin{table*}[t!]
    \centering
    \caption{Overview of the existing private inference methods, including the cryptographic schemes used, precision of neural networks supported, number of parties involved, evaluation setup configuration and availability of code.}
    \label{table:summaryexistingwork}
	\small
    \setlength\tabcolsep{3 pt}
    \begin{tabular}{@{}| l|c|c|c|c|c|c|c|@{}}
    \toprule
        \multicolumn{1}{|l|}{\multirow{2}{*}{\bf Prior Work}} & \multicolumn{1}{c|}{\multirow{2}{*}{\begin{tabular}[c]{@{}c@{}}\textbf{Cryptographic}\\\textbf{Scheme}\end{tabular}}} & \multicolumn{1}{c|}{\multirow{2}{*}{\textbf{Model Precision}}} & \multicolumn{1}{c|}{\multirow{2}{*}{ \textbf{Parties} }} & \multicolumn{3}{c|}{\textbf{Performance Evaluation Setup}} & \multicolumn{1}{c|}{\multirow{2}{*}{ \textbf{Code} }}  \\ 
        \cmidrule{5-7}
        \multicolumn{1}{|l|}{}  & \multicolumn{1}{c|}{}  & \multicolumn{1}{c|}{} & \multicolumn{1}{c|}{} & \multicolumn{1}{c|}{\textbf{CPU}} & \multicolumn{1}{c|}{\begin{tabular}[c]{@{}c@{}}\textbf{CPU Mark}\\\textbf{(Single Thread)\textsuperscript{f}}\end{tabular}} & \multicolumn{1}{c|}{\begin{tabular}[c]{@{}c@{}}\textbf{Relative }\\\textbf{CPU Mark}\end{tabular}} & \multicolumn{1}{c|}{} \\ 
        \midrule
        
        \textbf{MiniONN~\cite{MiniONN}}                     & \begin{tabular}[c]{@{}c@{}}Additively HE,\\ GC \end{tabular}      & Full                                                    & 2                                                       & \begin{tabular}[c]{@{}l@{}}Server: Intel Core i5 \\ 4 $\times$ 3.30 GHz cores\\ Client: Intel Core i5\\ 4 $\times$ 3.20 GHz cores\end{tabular} & 1,686-2,300                                                                                             & 0.61-0.83                                                                                            & Available\textsuperscript{a}                                                  \\ 
        \midrule
        
        \textbf{EzPC~\cite{EzPC}}                        & \begin{tabular}[c]{@{}c@{}}GC,\\ Additive SS \end{tabular}        & Full                                                    & 2                                                       & \begin{tabular}[c]{@{}c@{}}Intel Xeon E5-2673 v3\\2.40GHz\end{tabular}                                                                         & 1,723                                                                                                  & 0.62                                                                                                 & Available\textsuperscript{b}                                                   \\ 
        \midrule        
        
        \textbf{DeepSecure~\cite{DeepSecure}}                  & GC                                                                &  \begin{tabular}[l]{@{}l@{}}16-bit\\fixed-point\end{tabular}                                                       & 2                                                       & \begin{tabular}[c]{@{}l@{}}Intel Core i7-2600\\3.40GHz\end{tabular}                                                                            & 1,737                                                                                                  & 0.63                                                                                                 & --                                                    \\ 
       \midrule

        \textbf{SecureML~\cite{SecureML}}               & \begin{tabular}[c]{@{}c@{}}Linear HE,\\ GC \end{tabular}          & Full                                                    & 2                                                       & \begin{tabular}[c]{@{}c@{}}AWS c4.8xlarge \\ (Intel Xeon E5-2666 v3\\2.90 GHz) \end{tabular}                                                   & 1,918                                                                                                  & 0.69                                                                                                 & --                                                    \\ 
        \midrule
        
        \textbf{Gazelle~\cite{Gazelle}}                     & \begin{tabular}[c]{@{}c@{}}Additively HE,\\ GC \end{tabular}      & Full                                                    & 2                                                       & \begin{tabular}[c]{@{}c@{}}AWS c4.xlarge\\ (Intel Xeon E5-2666 v3\\2.90GHz)\end{tabular}                                                       & 1,918                                                                                                  & 0.69                                                                                                 & --                                                    \\ 
       \midrule

        \textbf{Delphi~\cite{Delphi}}                      & \begin{tabular}[c]{@{}c@{}}Additively HE,\\ GC \end{tabular}      & Full                                                    & 2                                                       & \begin{tabular}[c]{@{}c@{}}AWS c5.2xlarge\\ (Intel Xeon 8000 series\\3.0 GHz)\end{tabular}                                                     & 2,082                                                                                                  & 0.75                                                                                                 & Available\textsuperscript{c}                                                   \\ 
        \midrule
        
	\textbf{\method{}}                      & GC                                                                & Ternary                                                 & 2                                                       & \begin{tabular}[c]{@{}c@{}}AWS c5.2xlarge\\ (Intel Xeon 8124M\\3.0 GHz)\end{tabular}                                                           & 2,082                                                                                                  & 0.75                                                                                                 & Available\textsuperscript{d}                                                   \\
        \midrule

        \textbf{Chameleon~\cite{Chameleon}}                   & \begin{tabular}[c]{@{}c@{}}GC,\\ GMW,\\ Additive SS \end{tabular} & Full                                                    & 3\textsuperscript{e}                                                     & \begin{tabular}[c]{@{}l@{}}Intel Core i7-4790\\3.60GHz\end{tabular}                                                                            & 2,266                                                                                                  & 0.82                                                                                                 & --                                                    \\ 
        \midrule
        
        \textbf{XONN~\cite{XONN}}                        & 
        GC                                                                & Binary                                                  & 2                                                       & \begin{tabular}[c]{@{}l@{}}Intel Core i7-7700k\\4.5GHz\end{tabular}                                                                            & 2,777                                                                                                  & 1.0                                                                                                  & --                                                    \\ 
        \bottomrule

        \multicolumn{8}{C{15.5cm}}{\textsuperscript{a} MiniONN: \url{https://github.com/SSGAalto/minionn}}\\
        \multicolumn{8}{C{15.5cm}}{\textsuperscript{b} EzPC: \url{https://github.com/mpc-msri/EzPC}}\\
        \multicolumn{8}{C{15.5cm}}{\textsuperscript{c} Delphi: \url{https://github.com/mc2-project/delphi}}\\
        \multicolumn{8}{C{15.5cm}}{\textsuperscript{d} \method{}: \url{https://github.com/privacytrustlab/soteria_private_nn_inference}}\\
        \multicolumn{8}{C{15.5cm}}{\textsuperscript{e} Chameleon only uses the third party in pre-processing stage.}\\
        \multicolumn{8}{C{15.5cm}}{\textsuperscript{f} CPU mark: The configurations are listed with their the single-threaded CPU Mark scores as reported by \url{cpubenchmark.net/singleThread.html}. These single-thread benchmarks test processors on a variety of tasks, from floating point operations, string sorting and data compression (\url{https://www.cpubenchmark.net/cpu_test_info.html}) to provide an estimate of the capabilities of a processor. As microarchitectural optimizations vary from processor to processor, frequency alone cannot be used as a performance metric. The absolute scores and relative scores (compared to the highest scoring CPU in the table) for CPUs used in evaluation of related work are reported.}
    \end{tabular}
\end{table*}

\section{Empirical Evaluation} \label{sec:evaluation}

We evaluate the efficiency of our method in two ways.  We show how using ternary neural networks on  fixed model architectures, as used in the prior work, can reduce the overhead of secure inference on neural networks.  This is due to the sparsity of such models.  We also present the performance of \method{} architectures, in which model complexity is optimized along with the model accuracy.  

\subsection{Experimental Setup}

We evaluate our work on MNIST and CIFAR10 image classification datasets, as they have been extensively used in the literature to evaluate the performance of cryptographically secure neural network schemes.  We run our experiments on an AWS c5.2xlarge instance, running Ubuntu 18.04 LTS on an Intel Xeon 8124M at 3.0 GHz.

We use PyTorch 1.3~\cite{Pytorch_web}, a python-based deep learning framework to implement our architecture search algorithm and train the ternary models.  We use Synopsys Design Compiler~\cite{SynopsysDC_web}, version L-2016.03-SP5-2, to synthesize SystemVerilog code into the gate-level netlist. Our synthesis runs the TinyGarble gate library infrastructure\footnote{We use TinyGarble with  21ecca7cb75b33fd7508771fd35f03657dd44e5e gitid on \url{https://github.com/esonghori/TinyGarble} master branch.}.

We execute the garbled circuit protocol on the boolean circuit generated as described in previous sections.  We compute the number of non-XOR gates in the generated boolean circuit netlist as a measure of its complexity. We measure the exact performance of \method{} as its runtime during the offline and online phases of the protocol, and its communication cost. 

We present the experimental setup of  prior work and \method{}, including their CPU specification and link to available software codes, in Table~\ref{table:summaryexistingwork}.  We also present the details of all neural network architectures which we evaluate in this paper in Table~\ref{tab:architectures} in Appendix~\ref{Appendix:architectures}.

\subsection{Ternary Neural Networks}

As discussed in section~\ref{section:tnn}, we use ternary neural networks (TNNs) instead of binary networks as it  provides  significant performance gains with GC without any post-processing (i.e., the model is {\em trained} to be sparse). We perform two small experiments to illustrate the benefit of the sparsity (fraction of parameters with weight 0) of ternary models. Further, we analyze the effect of the scale of the network in the tradeoff between model accuracy and performance of private inference.

\begin{figure}[t!]
	\centering
	\begin{tikzpicture}

  \pgfplotsset{
      scale only axis,
      width=0.71\linewidth,
  }

  \begin{axis}[
    xticklabel style={
        /pgf/number format/fixed,
        /pgf/number format/precision=2
    },
    scaled y ticks=base 10:-3,
    xlabel= Sparsity (fraction of $0$-weight parameters),
    ylabel=Number of non-XOR gates,
    xmin=-0.02,
    xmax=0.35,
    xmajorgrids={true},
    ymajorgrids={true},
    xtick={0, 0.05, ..., 0.35},
    ymin=100000,
    ymax=270000,
    ytick distance=20000,
  ]
    \addplot[mark=*]
      coordinates{
        (0,258048)
        (0.11,196308)
        (0.22,172032)
        (0.33,122880)
      }; \label{plot_tnn_nonxor}  

  \end{axis}

\end{tikzpicture}
	\caption{Sparsity of a model versus its circuit complexity.  We measure sparsity as the fraction of model parameters with $0$ weight. We quantify circuit complexity as the number of non-XOR gates.  The numbers are computed on a ternary neural network with a $3\times3$ convolution operation with $4$ kernels and a $32\times32\times3$ input. For this experiment, we assign 0 weights to a random set of parameters, to get different levels of sparsity and corresponding number of non-XOR gates.}
	\label{fig:TNN-nonXOR}
\end{figure}
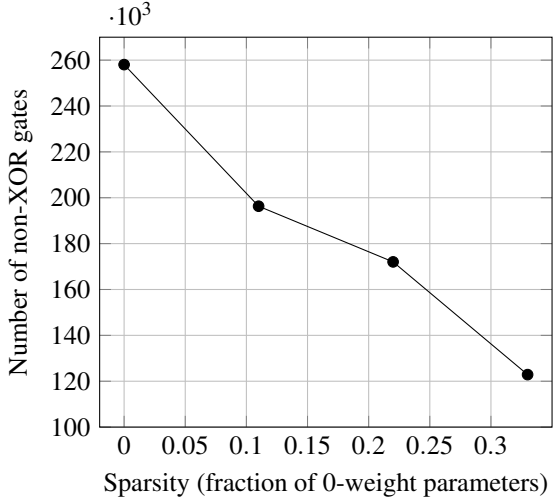

\paragraph{\em Sparsity.} 
Figure~\ref{fig:TNN-nonXOR} shows the number of non-XOR gates for a toy example: a 4-kernel $3\times3$ convolution operation taking input of size $32\times32\times3$ with padding sized $1$ ternary neural network.  We randomly set a fraction of parameters to zero to manually control the sparsity of the model. A BNN model is equivalent to the case where the sparsity is $0$.  We observe that as the sparsity increases the number of non-XOR gates decrease with almost the same factor. This can result in reducing both the communication overhead and the inference runtime, as we will see in training large models.

\begin{table}[t!]
    \centering
    \caption{Number of parameters and corresponding sparsity and trained model test accuracy for MNIST (m1) architecture with various levels of scaling factor. The table reports the statistics of the experiment in Figure~\ref{fig:scaling-factor}.}
    \label{tab:scaling-factor}
    \setlength\tabcolsep{3 pt}
\begin{tabular}{@{}ccccc @{}}
\toprule
    \multicolumn{1}{c}{\begin{tabular}[c]{@{}c@{}} \textbf{Scaling}\\\textbf{ Factor} \end{tabular}} & \multicolumn{1}{c}{\begin{tabular}[c]{@{}c@{}}\textbf{Total no. of}\\\textbf{ Parameters} \end{tabular}} & \multicolumn{1}{c}{\begin{tabular}[c]{@{}c@{}}\textbf{No. of}\\\textbf{ 0-weights} \end{tabular}} & \textbf{Sparsity} & \multicolumn{1}{c}{\textbf{Accuracy} }  \\ 
    \midrule
    0.25     & 26,432           & 5,946     & 0.22   & 0.9063 \\
    0.50     & 54,912           & 13,058    & 0.24   & 0.9413 \\
    0.75     & 85,440           & 18,703    & 0.22   & 0.9544 \\
    1.00     & 118,016          & 25,317    & 0.21   & 0.9579 \\
    1.50     & 189,312          & 43,428    & 0.23   & 0.9618 \\
    2.00     & 268,800          & 58,040    & 0.22   & 0.9658 \\
    2.50     & 356,480          & 83,451    & 0.23   & 0.9676 \\
    3.00     & 452,352          & 107,343   & 0.24   & 0.9712 \\
\end{tabular}
\end{table}

\begin{figure}[t!]
	\begin{tikzpicture}

  \pgfplotsset{
      scale only axis,
  }

  \begin{axis}[
    axis y line*=left,
    xlabel=Scaling Factor,
    ylabel=Test accuracy,
    xmin=0,
    xmax=3.2,
    xmajorgrids={true},
    xtick={0.5, 1.0, 1.5, 2.0, 2.5, 3.0},
    ymin=0.88,
    ymax=1.0,
  ]
    \addplot[color=blue,mark=*]
      coordinates{
        (0.25,0.9063)
        (0.50,0.9413)
        (0.75,0.9544)
        (1.00,0.9579)
        (1.50,0.9618)
        (2.00,0.9658)
        (2.50,0.9676)
        (3.00,0.9712)
      }; \label{plot_scalef_accuracy}

    \end{axis}

    \begin{axis}[
    xmin=0,
	xmax=3.2,    
      axis y line*=right,
      axis x line=none,
      ylabel=Inference runtime (ms),
      ymin=20.0,
      ymax=95.0,
      legend pos= north west,
    ]
    \addlegendimage{/pgfplots/refstyle=plot_scalef_accuracy}\addlegendentry{Accuracy}
    \addplot[color=red,mark=x]
      coordinates{
        (0.25,24.5)
        (0.50,36.1)
        (0.75,46.0)
        (1.00,53.4)
        (1.50,64.3)
        (2.00,74.5)
        (2.50,79.9)
        (3.00,90.1)
      }; \label{plot_scalef_runtime}
 
    \addlegendentry{Runtime}
  \end{axis}

\end{tikzpicture}
	\caption{Test accuracy versus inference runtime for a ternary neural network trained on a fixed MNIST (m1) architecture, in various scale. See Appendix~\ref{Appendix:architectures} for description of the model.}
	\label{fig:scaling-factor}
\end{figure}
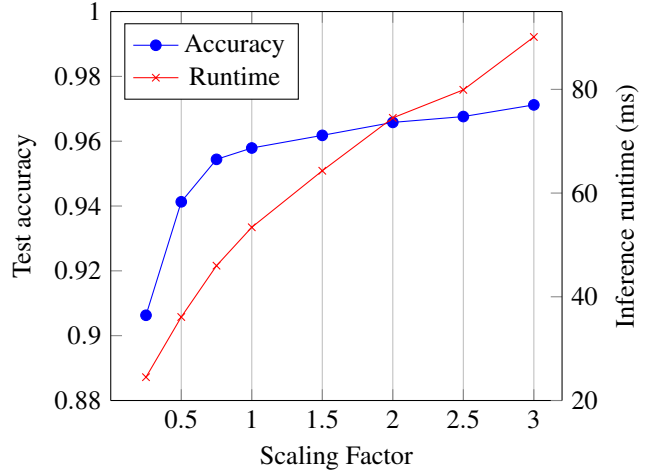

\begin{table*}[t]
    \centering
    \caption{Performance of private inference on garbled circuit models with binary versus ternary parameters. We show the independent offline and online runtimes, communication costs and number of non-XOR gates for three different types of operations. The operations are taken from MNIST (m3) network, trained in both binary (BNN) and ternary (TNN) configurations using a scaling factor of 1.}
    \label{tab:component-costs}
    \setlength{\tabcolsep}{3 pt}
    \small
\begin{tabular}{@{}|l|l|cc|cc|cc|cc|cc|c| @{}} 
    \toprule
    \multirow{3}{*}{\textbf {Operation  Type}} & \multirow{3}{*}{\textbf{Size} }    & \multicolumn{6}{c|}{\textbf{Runtime (ms)} }   & \multicolumn{2}{c|}{\multirow{2}{*}{\begin{tabular}[c]{@{}c@{}}\textbf{Communication}\\\textbf{(KB)} \end{tabular}}} & \multicolumn{2}{c|}{\multirow{2}{*}{\begin{tabular}[c]{@{}c@{}}\textbf{Number of}\\\textbf{non-XOR gates} \end{tabular}}} & \multirow{3}{*}{\begin{tabular}[c]{@{}c@{}}\textbf{TNN}\\\textbf{Sparsity}\end{tabular}}  \\ 
    \cmidrule{3-8}
    &    & \multicolumn{2}{c|}{\textbf{Offline} } & \multicolumn{2}{c|}{\textbf{Online} } & \multicolumn{2}{c|}{\textbf{Total} } & \multicolumn{2}{c|}{}    & \multicolumn{2}{c|}{} &  \\ 
    \cmidrule{3-12}
    &    & \textbf{BNN}  & \textbf{TNN}   & \textbf{BNN}  & \textbf{TNN}  & \textbf{BNN}  & \textbf{TNN} & \textbf{\hspace{0.15cm}BNN\hspace{0.15cm}}  & \textbf{TNN} & \textbf{BNN}  & \textbf{TNN}  &  \\ 
    \midrule
    Convolution    & \begin{tabular}[c]{@{}l@{}}Input: $12\times 12\times 16$\\ Padding: 0\\ Window: $5\times 5$\\ Kernels: 16\end{tabular} & 39.16 & 28.27  & 52.87 & 38.17 & 92.03 & 66.43    & 2,572 & 1,876    & 589,824   & 425,558   & 0.27 \\ 
    \midrule
    Maxpool    & \begin{tabular}[c]{@{}l@{}}Input:~$8\times 8\times 16$\\ Window:~$2\times 2$\end{tabular}  & \multicolumn{2}{c|}{0.35}  & \multicolumn{2}{c|}{0.57} & \multicolumn{2}{c|}{0.92}    & \multicolumn{2}{c|}{34}  & \multicolumn{2}{c|}{768}  & N.A. \\ 
    \midrule
    \begin{tabular}[c]{@{}l@{}} Fully\\Connected \end{tabular} & \begin{tabular}[c]{@{}l@{}}Input: 100\\ Nodes: 10 \end{tabular}    & 0.49  & 0.34   & 0.83  & 0.60  & 1.32  & 0.94 & 68    & 47   & 3150  & 2246  & 0.35 \\
    \bottomrule
\end{tabular}
\end{table*}

Note that while training a TNN, we cannot control the sparsity of the network. In Table~\ref{tab:component-costs}, we show the result of training binary and ternary models on MNIST dataset, model architecture m3. The table reports the GC costs of the components of the network for both BNNs and TNNs.  We can observe that the ternary model has a significant sparsity (about $0.3$).  This results in constructing smaller circuits for the model, which reduces the inference cost of using GC protocol over ternary neural networks.  This is reflected in the smaller number of non-XOR gates in the ternary circuits constructed on convolution and fully connected operations. The costs for maxpool operation will remain the same, as it does not have any learnable parameter.

\paragraph{\em Scale.}
As discussed in Section~\ref{sec:gcbinary}, we  need to scale the network operations to achieve a higher capacity for binary and ternary models and obtain better accuracies.  Figure~\ref{fig:scaling-factor} demonstrates the impact of scale of the network on accuracy of the model and its GC runtime for private inference. Although inference time increases linearly with the scaling factor, accuracy improves upto a certain extent with scaling (scaling factor of 1) and then becomes almost constant. This denotes that we can select a sweet spot for scaling factor thereby optimizing for both accuracy and performance.

Table~\ref{tab:scaling-factor} shows how scaling factor affects the number of parameters in the network.  As the scaling factor in TNNs increases, the accuracy increases. However, with diminishing returns after a certain limit. The inference cost of the circuit also increases, as is evident from the growth of runtime with change in scaling factor.  Note that the sparsity is about $0.24$ for scaling factor $3$ for the ternary neural network, which means that the effective size of the model (hence its performance cost) remains comparable to a binary neural network (without any scaling), albiet with better accuracy. As a reference, the test accuracy of a BNN model with the same architecture is $0.9514$.

\paragraph{\em Comparison with prior work.}
Table~\ref{tab:comparison-results} reports the results of  our experiments when compared to prior work.  In this subsection, we present the outcome of basic \method{} on {\em fixed} model architectures which are used in the literature, thus only discussing the effect of sparsity of ternary neural networks on the tradeoff between accuracy and performance costs.
We use three different architectures for each of the two datasets, which have been used in existing work. m1-3 are used with the MNIST dataset, while m4-6 are used with the CIFAR10 dataset.  See Appendix~\ref{Appendix:architectures} for the descriptions of model architectures.  We use the same scaling factors for our networks as used by XONN~\cite{XONN}, which is the only other comparable work with quantized (binary) weights and inputs, for a fair comparison.

We observe that for MNIST datasets, the basic TNN \method{} models (m1-m3) provide better  runtime and communication performance on average than prior work with maximum drop in accuracy of only $0.016$ (for model m3). This shows that \method{} is useful in designing custom models that provide optimal performance guarantees while retaining high prediction accuracy. For CIFAR10 datasets, we observe that for models used in prior work (m4 to m6), our basic TNN models exhibit a slightly higher drop in accuracy of $0.8$, but provide a computation and communication gain, on average, as compared to prior work. Overall, our results show that \method{} provides a flexible approach of training private models given the constraint on performance and accuracy of the model.

\begin{table*}[t!]
    \centering
    \caption{
        Performance analysis of existing secure schemes for private neural network inference. We compare \method{} constructed on fixed model architectures with ternary parameters, as well as optimal architectures constructed by \method{}, with the prior work. We provide the descriptions of the model architectures in Appendix~\ref{Appendix:architectures}).  We use the same scaling factor for \method{} and XONN for fixed model architectures (1.75 for m1, 4.0 for m2, 2.0 for m3, 2.0 for m4, 3.0 for m5, 2.0 for m6), and use 3.0 for MNIST (\method) and CIFAR10 (\method) for the models constructed by our architecture search algorithm.} 
    \label{tab:comparison-results} 
    \begin{tabular}{@{} |lld{3.2}d{3.2}d{3.2}d{5.2}l| @{}}
        \toprule
        \textbf{Model} & \textbf{Secure Scheme} & \multicolumn{3}{c}{\textbf{Runtime (s)}} & 
        \multicolumn{1}{c}{\textbf{Communication (MB)}} & \textbf{Test Accuracy} \\ \cmidrule{3-5}
        &  & \multicolumn{1}{c}{\textbf{Offline}} & \multicolumn{1}{c}{\textbf{Online}} & \multicolumn{1}{c}{\textbf{Total}} &  &  \\ \midrule
        
        \multirow{6}{*}{MNIST (m1)} & SecureML~\cite{SecureML} & 4.7 & 0.18 & 4.88 & -\textsuperscript{d} & 0.931 \\
        & MiniONN~\cite{MiniONN} & 0.9 & 0.14 & 1.04 & 15.8 & 0.976 \\
        & EzPC~\cite{EzPC} & - & -\textsuperscript{c} & 0.7 & 76 & 0.976 \\
        & Gazelle~\cite{Gazelle} & 0 & 0.03 & 0.03 & 0.5 & 0.976 \\
        & XONN~\cite{XONN} & - & -\textsuperscript{c} & 0.13\textsuperscript{b} & 4.29 & 0.976\textsuperscript{a}\hspace{0.3cm}(0.9591) \\
        & \method{}  (TNN) & 0.04 & 0.03 & 0.07 & 3.72 & 0.9642 \\ \midrule
        
        \multirow{6}{*}{MNIST (m2)} & DeepSecure~\cite{DeepSecure} & 7.69 & 1.98 & 9.67 & 791 & 0.9895 \\
        & MiniONN & 0.88 & 0.4 & 1.28 & 47.6 & 0.9895 \\
        & EzPC & - & -\textsuperscript{c} & 0.6 & 70 & 0.990 \\
        & Gazelle & 0.15 & 0.05 & 0.20 & 8.0 & 0.990 \\
        & XONN & - & -\textsuperscript{c} & 0.16\textsuperscript{b} & 38.28 & 0.9864\textsuperscript{a}\hspace{0.18cm}(0.9718) \\
        & \method{}  (TNN) & 0.08 & 0.06 & 0.14 & 30.68 & 0.9733 \\ \midrule
        
        \multirow{5}{*}{MNIST (m3)} & MiniONN & 3.58 & 5.74 & 9.32 & 657.5 & 0.990 \\
        & EzPC & - & -\textsuperscript{c} & 5.1 & 501 & 0.990 \\
        & Gazelle & 0.48 & 0.33 & 0.81 & 70 & 0.990 \\
        & XONN & - & -\textsuperscript{c} & 0.15\textsuperscript{b} & 32.13 & 0.990\textsuperscript{a}\hspace{0.3cm}(0.9672) \\
        & \method{}  (TNN) & 0.08 & 0.07 & 0.15 & 26.04 & 0.9740 \\ \midrule
        
        \multirow{2}{*}{MNIST (\method{})}
        & \method{}  ($\lambda=0.6$) & 0.09 & 0.08 & 0.17 & 36.24 & 0.9883 \\
        & \method{}  ($\lambda=0$) & 0.016 & 0.018 & 0.034 & 95.18 & 0.9811 \\ \midrule \midrule 
        
        \multirow{2}{*}{CIFAR10 (m4)} & XONN & - & -\textsuperscript{c} & 15.07\textsuperscript{b} & 4980 & 0.80\textsuperscript{a}\hspace{0.47cm}(0.7197) \\
        & \method{}  (TNN) & 8.56 & 6.14 & 14.70 & 936.1 & 0.7314 \\ \midrule
        
        \multirow{6}{*}{CIFAR10 (m5)} & MiniONN & 472 & 72 & 544 & 9272 & 0.8161 \\
        & EzPC & - & -\textsuperscript{c} & 265.6 & 40683 & 0.8161 \\
        & Gazelle & 9.34 & 3.56 & 12.9 & 1236 & 0.8161 \\
        & Delphi\textsuperscript{e} & 45 & 1 & 46 & 200 & 0.85 \\
        & XONN & - & -\textsuperscript{c} & 5.79\textsuperscript{b} & 2599 & 0.8185\textsuperscript{a}\hspace{0.18cm}(0.7266) \\
        & \method{}  (TNN) & 3.48 & 2.95 & 6.43 & 461.3 & 0.7252 \\ \midrule
        
        \multirow{2}{*}{CIFAR10 (m6)} & XONN & - & -\textsuperscript{c} & 16.09\textsuperscript{b} & 5320 & 0.83\textsuperscript{a}\hspace{0.50cm}(0.7341) \\
        & \method{}  (TNN) & 9.01 & 6.63 & 15.64 & 982.7 & 0.7396 \\ \midrule
        
        \multirow{2}{*}{CIFAR10 (\method{})}
        & \method{}  ($\lambda = 0.6$) & 3.69 & 3.26 & 6.95 & 497.6 & 0.7384 \\
        & \method{}  ($\lambda = 0$) & 4.01 & 3.53 & 7.54 & 561.2 & 0.7211 \\ \bottomrule
        \multicolumn{7}{C{15.5cm}}{\textsuperscript{a}We could not reproduce the test accuracies for XONN. We report the results that we obtained on the same model architectures with the same setting in the respective paper in parenthesis.}\\
        \multicolumn{7}{C{15.5cm}}{\textsuperscript{b}XONN's runtime reported by the authors is measured on a high-performance Intel processor, which is faster than the one used by all other methods.}\\
        \multicolumn{7}{C{15.5cm}}{\textsuperscript{c}Breakdown of runtime cost into offline and online runtime is not reported by the authors.}\\
        \multicolumn{7}{C{15.5cm}}{\textsuperscript{d}Communication cost is not reported by the authors.}\\
        \multicolumn{7}{C{15.5cm}}{\textsuperscript{e}The runtime, communication cost and test accuracy are visual estimates from the graphs reported by the authors.}
    \end{tabular}
    
\end{table*}

\subsection{Architecture Search}

We present the details of our empirical analysis of \method{}. In particular, we evaluate the cost function we used in the architecture search algorithm, the effect of the performance regularization during the search, the effect of the model size on the tradeoff between accuracy and inference runtime, and compare \method{} with the prior work.

\paragraph{\em Implementation.} 
To handle conversion of the model into a digital circuit supported by TinyGarble, we first build a representation of the model and parameters in SystemVerilog, and then synthesize and optimize our circuit (using Synopsys Design Compiler) to use circuit elements supported by TinyGarble. In this first step, we designed a collection of parameterized components (notably dot product, and maxpool) to use as building blocks our architecture search algorithm. Each component is flexibly designed to efficiently accept arbitrary size input and output, and is composed to form the complete model. In a general setting, hardware-level code is typically straight-forward to generate. However, to enable  TNNs with \method{}, we have to dynamically define the sparsity of the modules depending on the result of model training and architecture search. Along with the parameter data, we define the sparsity information which is used during \texttt{generate} phases in SystemVerilog to build the sparse network in hardware (taking advantage of the 0-valued parameters of the model). Altogether, this allows us to build and evaluate the models constructed by \method{}.

\begin{table}[t!]
    \centering
    \caption{Runtime and communication cost of each operation based on their garbled circuit inference. We also calculate the performance penalty factor (which we use in our regularized architecture search algorithm) as the average of the relative costs for each unit w.r.t the most expensive operation.}
    \label{table:operationcost}
    \setlength{\tabcolsep}{6pt}
\begin{tabular}{@{}l |ccc@{}}
\toprule
    \textbf{Operation} & \textbf{\begin{tabular}[c]{@{}c@{}}Runtime\\ (ms)\end{tabular}} & \textbf{\begin{tabular}[c]{@{}c@{}}Comm.\\ (KB)\end{tabular}} & \textbf{\begin{tabular}[c]{@{}c@{}}Penalty\\ factor\end{tabular}} \\ \midrule
    {CONV$5\times5$} & 55.40 & 7942 & 1.00 \\
    {CONV$3\times3$} & 23.10 & 3190 & 0.41 \\
    {MAXPOOL$2\times2$} & 3.23 & 145 & 0.04 \\
    {IDENTITY} & 0.00 & 0.00 & 0.00
\end{tabular}
\end{table}

\paragraph{\em Cost function for regularized architecture search.}
As presented in Section~\ref{sec:method}, our algorithm searches for the models that are not only accurate but also are efficient with respect to the costs of using garbled circuits on the model architecture.  For this, we modify the score value that the DARTS architecture search algorithm gives to each operation (e.g., maxpool, or convolution with different dimensions) with a regularized penalty factor proportional to the performance cost of the operation.  Table~\ref{table:operationcost} presents the communication and runtime cost of each operation that we use in our algorithm.  The penalty factor is computed as the average of the relative communication cost and relative runtime cost of each operation, with respect to the most costly operation ({CONV$5\times5$}). We use this penalty factor in the experiments.

\begin{table}[t!]
\centering
\caption{Number of parameters and corresponding sparsity and test accuracy with different levels of $\lambda$ for \method{} architecture search over MNIST dataset with 1 cell architecture having 4 sequential operations. The setting is the same as the one illustrated in Figure~\ref{fig:lambda-vs-accuracy}(b). The scaling factor is 3. For larger $\lambda$, the algorithm penalizes constructing large models.}
\label{tab:lambda-parameters}
\small
\begin{tabular}{@{}ccccc@{}}
\toprule
    \multicolumn{1}{c}{\textbf{$\lambda$}} & \multicolumn{1}{c}{\begin{tabular}[c]{@{}c@{}}\textbf{Total no. of}\\\textbf{Parameters}\end{tabular}} & \multicolumn{1}{c}{\begin{tabular}[c]{@{}c@{}}\textbf{No. of}\\\textbf{0-weights}\end{tabular}} & \multicolumn{1}{c}{\textbf{Sparsity}} & \multicolumn{1}{c}{\textbf{Accuracy}}  \\ 
    \midrule
    1.0      & 133,032      & 41,212               & 0.31    & 0.8892   \\
    0.8      & 729,768      & 200,540              & 0.27    & 0.9721   \\
    0.6      & 2,904,168    & 1,080,217            & 0.37    & 0.9883   \\
    0.4      & 2,904,168    & 1,080,217            & 0.37    & 0.9883   \\
    0.2      & 2,941,032    & 895,034              & 0.30    & 0.9887   \\
    0.0      & 11,466,600   & 5,116,030            & 0.45    & 0.9811  
\end{tabular}
\end{table}

\begin{figure}[t!]
	\centering
	\begin{tikzpicture}

  \pgfplotsset{
      scale only axis,
      width=0.71\linewidth,
  }

  \begin{axis}[
    xlabel=Inference runtime (s),
    ylabel=Test accuracy,
    xmin=1,
    xmax=8,
    xmajorgrids={true},
    ymajorgrids={true},
    xtick={0, 1, ..., 8},
    ymin=0.1,
    ymax=0.8,
  ]
    \addplot[mark=*]
      coordinates{
        (7.5,0.72)
        (6.7,0.74)
        (6.7,0.74)
        (5.7,0.73)
        (4.2,0.45)
        (1.6,0.17)
      }; \label{plot_lambda_accuracy}  
  \node [above,color=blue] at (axis cs: 7.5, 0.73) {$0$};
  \node [above,color=blue] at (axis cs: 6.7, 0.66) {$0.2$};
  \node [above,color=blue] at (axis cs: 6.7, 0.60) {$0.4$};
  \node [above,color=blue] at (axis cs: 5.7, 0.74) {$0.6$};
  \node [above,color=blue] at (axis cs: 4.2, 0.35) {$0.8$};
  \node [above,color=blue] at (axis cs: 2, 0.10) {$\lambda=1.0$};

  \end{axis}

\end{tikzpicture}
	\\(a) CIFAR10
	\\[5pt]
	\begin{tikzpicture}

  \pgfplotsset{
      scale only axis,
      width=0.71\linewidth,
  }

  \begin{axis}[
    xlabel=Inference runtime (s),
    ylabel=Test accuracy,
    xmin=0,
    xmax=0.4,
    xmajorgrids={true},
    ymajorgrids={true},
    xtick={0, 0.1, 0.2, 0.3, 0.4},
    ymin=0.85,
    ymax=1,
  ]
    \addplot[mark=*]
      coordinates{
        (0.34,0.9811)
        (0.23,0.9887)
        (0.17,0.9883)
        (0.17,0.9883)
        (0.12,0.9721)
        (0.04,0.8892)
      };
  \node [above,color=blue] at (axis cs: 0.34, 0.984) {$0$};
  \node [above,color=blue] at (axis cs: 0.23, 0.989) {$0.2$};
  \node [above,color=blue] at (axis cs: 0.17, 0.989) {$0.4$};
  \node [above,color=blue] at (axis cs: 0.17, 0.974) {$0.6$};
  \node [above,color=blue] at (axis cs: 0.096, 0.967) {$0.8$};
  \node [above,color=blue] at (axis cs: 0.04, 0.875) {$\lambda=1.0$};

  \end{axis}

\end{tikzpicture}
	\\(b) MNIST
	\caption{Inference runtime versus test accuracy of a garbled ternary model, constructed with \method architecture search algorithm, for various values of circuit cost regularization $\lambda$.  We obtain the architecture for a neural network with (a) $3$ cells for CIFAR10 dataset, and (b) $1$ cell for MNIST dataset. All experiments use a scaling factor of 3.}
	\label{fig:lambda-vs-accuracy}
\end{figure}

\paragraph{\em Balancing accuracy and inference costs.}
For the architecture search in \method{}, we balance accuracy and inference cost over GC protocol, using a regularization factor $\lambda$. With $\lambda=1$ the importance of the penalty factor is maximum, and $\lambda=0$ represents the case where we ignore the performance cost. We execute the search process with different values of lambda.  We constrain the search to finding a 3-cell architecture for CIFAR10 dataset and 1-cell architecture for MNIST dataset, with each cell having 4 operations in sequence. We run the architecture search algorithm for 100 epochs, and subsequently train the obtained architectures for 200 epochs.

Figure~\ref{fig:lambda-vs-accuracy} presents trade-off between test accuracy of the optimal architectures and their inference runtime.  Table~\ref{tab:lambda-parameters} provides the statistics on the number of model parameters and the model sparsity for different values of $\lambda$ for the MNIST dataset. As $\lambda$ increases, cheaper operations, that have fewer trainable parameters are chosen by the search process, which improves the inference runtime at the expense of the model accuracy.  As we observe, $\lambda=0.6$ can provide a reasonable balance between both accuracy and the inference cost.  This is where the search algorithm identifies cheaper operations that collectively can result in equivalent accuracy that can be achieved using more expensive operations. It is very important to note that selecting $\lambda$ depends on how much cost or accuracy drop we can tolerate for a given setting. Thus, \method{} enables adapting private inference to the specific requirements and limitations of a system.

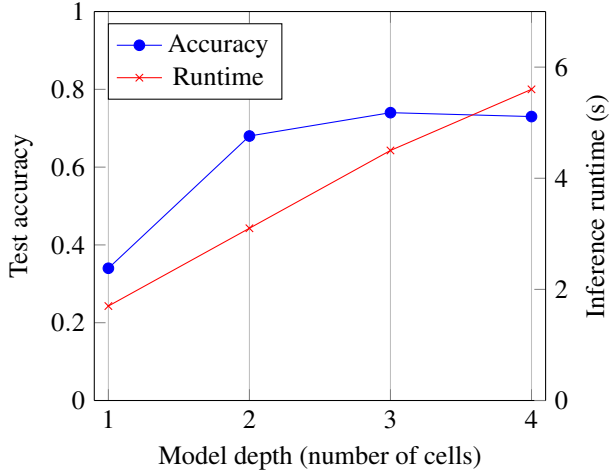
\begin{figure}[t!]
	\begin{tikzpicture}

  \pgfplotsset{
      scale only axis,
  }

  \begin{axis}[
    axis y line*=left,
    xlabel= Model depth (number of cells),
    ylabel=Test accuracy,
    xmin=0.9,
    xmax=4.1,
    xmajorgrids={true},
    xtick={1,2,...,4},
    ymin=0.0,
    ymax=1.0,
  ]
    \addplot[color=blue,mark=*]
      coordinates{
        (1,0.34)
        (2,0.68)
        (3,0.74)
        (4,0.73)
      }; \label{plot_numcells_accuracy}

    \end{axis}

    \begin{axis}[
        xmin=0.9,
    xmax=4.1,
      axis y line*=right,
      axis x line=none,
      ylabel= Inference runtime (s),
      ymin=0.0,
      ymax=7.0,
      legend pos= north west,
    ]
    \addlegendimage{/pgfplots/refstyle=plot_numcells_accuracy}\addlegendentry{Accuracy}
    \addplot[color=red,mark=x]
      coordinates{
        (1,1.7)
        (2,3.1)
        (3,4.5)
        (4,5.6)
      }; \label{plot_numcells_runtime}

    \addlegendentry{Runtime}
  \end{axis}

\end{tikzpicture}
	\caption{Impact of the model's depth (as the number of cells in the \method{} architecture search algorithm) on test accuracy and inference runtime of garbled ternary model on CIFAR10. Regularization term $\lambda$ is $0.6$. Scaling factor is 3.}
	\label{fig:numcells-vs-accuracy}
\end{figure}

\paragraph{\em Finding the optimal depth for the model (number of cells).}
The number of cells that the final architecture will have is a manually-set hyperparameter, that practically defines the depth of the model architecture. We perform an experiment on the CIFAR10 dataset, using cells with 4 operations in sequence and $\lambda=0.6$. We run the search process for 100 epochs, and train each resultant architecture for 200 epochs. Figure~\ref{fig:numcells-vs-accuracy} presents the results of executing the search with different number of cells.  It illustrates the model accuracies along with their inference runtime.

We can observe that for a single cell architecture, the accuracy levels are low, as the network could not process the features required to perform a generalizable classification. The test accuracy peaks for 3-cell architecture, suggesting that it has enough operations to process the required features of the inputs. In parallel, we can see that as the number of cells increases, the runtime also increases.

\paragraph{\em Comparison with prior work.}
Table~\ref{tab:comparison-results} reports the results for \method{} trained models that are optimized for both accuracy and efficiency. For the MNIST dataset, we observe that our model with $\lambda =0$ gives the best model as compared to prior work while balancing the runtime performance ($0.034$) and accuracy of $0.9811$. Increasing  $\lambda =0.6$ increases the accuracy by a small value.  Similarly, for CIFAR10 datasets, we observe that a \method{}-trained model with $\lambda =0.6$ gives better accuracy than prior work as compared to the numbers reported in brackets. In addition, \method{} models provide acceptable runtime performance and communication overhead that outperforms the results from prior work.  Our evaluation on MNIST and CIFAR10 datasets confirm that  \method{} is effective in training models that are customized to perform well for both performance and accuracy.

\section{Related Work}
There has been several approaches that introduce new techniques for secure machine learning, or build up on existing techniques by trying to optimize bottlenecks.

\paragraph{\em Homomorphic Encryption.} In CryptoNets~\cite{Cryptonets}\cite{Crypto-nets}, the authors modify the neural network operation by using square function as an activation and average pool instead of maxpool to reduce the non-linear functions to low degree polynomial to control the noise. Similar approaches of using homomorphic encryption on data and optimizing the machine learning operations to limit the noise have been explored extensively~\cite{HomomorphicDiscreteNN, MLConfidential, BostClassificationEncrypted}. 

Hesamifard et al.~\cite{PrivacyML_Cloud} explore using homomorphic encrypted data for training the neural networks. CryptoDL~\cite{CryptoDL} explores various activation functions with low polynomial degree that can work well with homomorphic encrypted data and proposed an activation using the derivative of ReLU function. However, using homomorphic encryption adds to an additional computational overhead and most of the non-linear activations cannot be effectively computed which results in a degradation in reliability of the deep learning systems.

\paragraph{\em Secure Multiparty Computation.} Secure multiparty computation requires a very low computation overhead but requires extensive communication between the parties. It has been used for several machine learning operations. 
DeepSecure~\cite{DeepSecure} only uses GC to compute all the operations in the neural network. They rely on pre-processing of the data by reducing the dimensions to improve the performance and is implemented on the TinyGarble~\cite{TinyGarble_paper} library. Chameleon~\cite{Chameleon} uses a combination of arithmetic sharing, garbled circuit and boolean sharing to compute the neural networks for secure inference. They rely on third party server to perform computation in the offline phase resulting in better performance than the previous work. 

XONN~\cite{XONN} leverages Binary Neural Networks with GC. Binarization dramatically reduces the inference latency for the network compared to other frameworks that utilize full-precision weights and inputs, as it converts matrix multiplications into simple XNOR-popcounts. They use TinyGarble library as well to implement the Boolean circuits for GC. 
Prio~\cite{Prio} uses a secret sharing~\cite{ShamirSS} based protocol to compute aggregate statistics over private data from multiple sources. They deploy a secret-shared non-interactive zero-knowledge proof mechanism to verify whether data sent by clients is well-formed, and then decode summed encodings of clients' data to generate aggregate statistic. They extend the application of Prio to foundational machine learning techniques such as least squares regression.

\paragraph{\em Hybrid Schemes.} A judicious combination of homomorphic encryption and multiparty computation protocol have shown to give some additional benefits in terms of runtime and communication costs. 
Gazelle~\cite{Gazelle} uses lattice based Packed Additive homomorphic encryption to compute dot product and convolution but relies on garbled circuits for implementing non-linear operations like Maxpool and ReLU. They reduce the overall bandwidth by packing ciphertexts and re-encryption to refresh the noise budget. Delphi~\cite{Delphi} builds upon this work and uses Architecture Search to select optimal replacement positions for expensive ReLU activation function with a quadratic approximation with minimal loss in accuracy. 

MiniONN~\cite{MiniONN} pre-computes multiplication triplets using homomorphic encryption for GMW protocol followed by SPDZ~\cite{MPC_SomewhatHE, Breaking_SPDZ_Limits} protocol. The multiplication triplets are exchanged securely using additive homomorphic encryption like Paillier or DGK.
SecureML~\cite{SecureML} uses garbled circuits and additive homomorphic encryption to speed up some NN operations. However, the conversion costs between of homomorphic encryption and Yao's garbled circuits is expensive and the performance of homomorphic encryption scales poorly with increasing security parameter~\cite{ABY}. 

Hence, we rely on only garbled circuit protocol to efficiently compute neural network operations during inference with low communication bandwidth, low computation complexity and low memory footprint using binary neural networks while maintaining the accuracy. Most of the previous work have relied heavily on optimizing the complex cryptographic operations to work well with the neural networks. We show that it is possible to optimize the neural network to get an efficient privacy preserving neural network architectures.

\paragraph{\em Trusted Computing.} Some research uses trusted processors where they assume that the underlying hardware is trustworthy and outsource all the machine learning computations to the trusted hardware. Chiron~\cite{hunt2018chiron} is a training system for privacy-preserving machine learning as a service which conceals the training data from the operator. It uses Intel Software Guard Extensions (SGX) and runs the standard ML training in an enclave and confines it in a Ryoan sandbox~\cite{Ryoan} to prevent it from leaking the training data outside the enclave.

Ohrimenko et al.~\cite{ohrimenko2016oblivious} propose a solution for secure multiparty ML by using trusted Intel SGX-enabled processors and used oblivious protocols between client and server where the input and outputs are blinded. However, the memory of enclaves is limited and it is difficult to process memory and computationally intensive operations like matrix multiplication in the enclaves with paralellism. To address this, Slalom~\cite{Slalom} provides a methodology to outsource the matrix multiplication to a faster untrusted processor and verify the computation.

\section{Conclusions}

We introduce \method{}, a system that takes advantage of the power of neural architecture search algorithms in order to design model architectures which jointly optimize accuracy and efficiency for private inference.  We use garbled circuits (GC) as our cryptographic primitive, due to its flexibility. However, other secure multi-party computation schemes can also be used to enrich the set of secure operations that could be chosen by the architecture search algorithm. .Instead of model post-processing, we also enable the stochastic gradient descent algorithm to train a sparse model (setting some parameters to 0), hence further improving the efficiency of the trained model.  To this end, we train ternary neural networks which have shown to have a significant potential in learning reasonably accurate models.  We construct optimal architectures that balance accuracy and inference efficiency on GC on MNIST and CIFAR10 datasets.  As opposed to the prior work that build cryptographic schemes around given fixed models, \method{} provides a flexible solution that can be adapted to the accuracy and performance requirements of any given system, and enables trading off between requirements.

\section*{Reproducibility}

The code for our work is available at \url{https://github.com/privacytrustlab/soteria_private_nn_inference}.


\begin{thebibliography}{10}
	
	\bibitem{Pytorch_web}
	Pytorch 1.3.
	\newblock https://pytorch.org/.
	
	\bibitem{SynopsysDC_web}
	Synopsys design compiler, version {L-2016.03-SP5-2}.
	\newblock
	https://www.synopsys.com/implementation-and-signoff/rtl-synthesis-test/dc-ultra.html.
	
	\bibitem{beaver1991efficient}
	Donald Beaver.
	\newblock Efficient multiparty protocols using circuit randomization.
	\newblock In Joan Feigenbaum, editor, {\em Advances in Cryptology --- CRYPTO
		'91}, pages 420--432, Berlin, Heidelberg, 1992. Springer Berlin Heidelberg.
	
	\bibitem{BostClassificationEncrypted}
	Raphael Bost, Raluca~Ada Popa, Stephen Tu, and Shafi Goldwasser.
	\newblock Machine learning classification over encrypted data.
	\newblock Cryptology ePrint Archive, Report 2014/331, 2014.
	\newblock \url{https://eprint.iacr.org/2014/331}.
	
	\bibitem{HomomorphicDiscreteNN}
	Florian Bourse, Michele Minelli, Matthias Minihold, and Pascal Paillier.
	\newblock Fast homomorphic evaluation of deep discretized neural networks.
	\newblock Cryptology ePrint Archive, Report 2017/1114, 2017.
	\newblock \url{https://eprint.iacr.org/2017/1114}.
	
	\bibitem{brakerski2011fully}
	Zvika Brakerski, Craig Gentry, and Vinod Vaikuntanathan.
	\newblock Fully homomorphic encryption without bootstrapping.
	\newblock Cryptology ePrint Archive, Report 2011/277, 2011.
	\newblock \url{https://eprint.iacr.org/2011/277}.
	
	\bibitem{EzPC}
	Nishanth Chandran, Divya Gupta, Aseem Rastogi, Rahul Sharma, and Shardul
	Tripathi.
	\newblock {EzPC}: Programmable, efficient, and scalable secure two-party
	computation for machine learning.
	\newblock In {\em IEEE European Symposium on Security and Privacy}, February
	2019.
	
	\bibitem{Prio}
	Henry Corrigan-Gibbs and Dan Boneh.
	\newblock Prio: Private, robust, and scalable computation of aggregate
	statistics.
	\newblock In {\em 14th {USENIX} Symposium on Networked Systems Design and
		Implementation ({NSDI} 17)}, pages 259--282, Boston, MA, March 2017. {USENIX}
	Association.
	
	\bibitem{MPC_SomewhatHE}
	I.~Damgard, V.~Pastro, N.P. Smart, and S.~Zakarias.
	\newblock Multiparty computation from somewhat homomorphic encryption.
	\newblock Cryptology ePrint Archive, Report 2011/535, 2011.
	\newblock \url{https://eprint.iacr.org/2011/535}.
	
	\bibitem{Breaking_SPDZ_Limits}
	Ivan Damgard, Marcel Keller, Enrique Larraia, Valerio Pastro, Peter Scholl, and
	Nigel~P. Smart.
	\newblock Practical covertly secure mpc for dishonest majority – or: Breaking
	the spdz limits.
	\newblock Cryptology ePrint Archive, Report 2012/642, 2012.
	\newblock \url{https://eprint.iacr.org/2012/642}.
	
	\bibitem{ABY}
	Daniel Demmler, Thomas Schneider, and Michael Zohner.
	\newblock {ABY} - {A} framework for efficient mixed-protocol secure two-party
	computation.
	\newblock In {\em 22nd Annual Network and Distributed System Security
		Symposium, {NDSS} 2015, San Diego, California, USA, February 8-11, 2015}. The
	Internet Society, 2015.
	
	\bibitem{Cryptonets}
	Nathan Dowlin, Ran Gilad-Bachrach, Kim Laine, Kristin Lauter, Michael Naehrig,
	and John Wernsing.
	\newblock {CryptoNets}: Applying neural networks to encrypted data with high
	throughput and accuracy.
	\newblock In {\em Proceedings of the 33rd International Conference on
		International Conference on Machine Learning - Volume 48}, ICML'16, pages
	201--210. JMLR.org, 2016.
	
	\bibitem{elgamal1985public}
	Taher ElGamal.
	\newblock A public key cryptosystem and a signature scheme based on discrete
	logarithms.
	\newblock In George~Robert Blakley and David Chaum, editors, {\em Advances in
		Cryptology}, pages 10--18, Berlin, Heidelberg, 1985. Springer Berlin
	Heidelberg.
	
	\bibitem{NAS_survey}
	Thomas Elsken, Jan~Hendrik Metzen, and Frank Hutter.
	\newblock Neural architecture search: A survey, 2018.
	
	\bibitem{FHE}
	Craig Gentry.
	\newblock {\em A Fully Homomorphic Encryption Scheme}.
	\newblock PhD thesis, Stanford, CA, USA, 2009.
	\newblock AAI3382729, Advisor: D. Boneh.
	
	\bibitem{gentry}
	Craig Gentry.
	\newblock Fully homomorphic encryption using ideal lattices.
	\newblock In {\em Proceedings of the Forty-First Annual ACM Symposium on Theory
		of Computing}, STOC ’09, page 169–178, New York, NY, USA, 2009.
	Association for Computing Machinery.
	
	\bibitem{GMW}
	O.~Goldreich, S.~Micali, and A.~Wigderson.
	\newblock {How to Play ANY Mental Game}.
	\newblock In {\em Proceedings of the Nineteenth Annual ACM Symposium on Theory
		of Computing}, STOC '87, pages 218--229, New York, NY, USA, 1987. ACM.
	
	\bibitem{MLConfidential}
	Thore Graepel, Kristin Lauter, and Michael Naehrig.
	\newblock {ML Confidential}: Machine learning on encrypted data.
	\newblock Cryptology ePrint Archive, Report 2012/323, 2012.
	\newblock \url{https://eprint.iacr.org/2012/323}.
	
	\bibitem{CryptoDL}
	Ehsan Hesamifard, Hassan Takabi, and Mehdi Ghasemi.
	\newblock {CryptoDL}: Deep neural networks over encrypted data, 2017.
	
	\bibitem{PrivacyML_Cloud}
	Ehsan Hesamifard, Hassan Takabi, Mehdi Ghasemi, and Catherine Jones.
	\newblock Privacy-preserving machine learning in cloud.
	\newblock In {\em Proceedings of the 2017 on Cloud Computing Security
		Workshop}, CCSW ’17, page 39–43, New York, NY, USA, 2017. Association for
	Computing Machinery.
	
	\bibitem{BNN}
	Itay Hubara, Matthieu Courbariaux, Daniel Soudry, Ran El-Yaniv, and Yoshua
	Bengio.
	\newblock Binarized neural networks.
	\newblock In D.~D. Lee, M.~Sugiyama, U.~V. Luxburg, I.~Guyon, and R.~Garnett,
	editors, {\em Advances in Neural Information Processing Systems 29}, pages
	4107--4115. Curran Associates, Inc., 2016.
	
	\bibitem{hunt2018chiron}
	Tyler Hunt, Congzheng Song, Reza Shokri, Vitaly Shmatikov, and Emmett Witchel.
	\newblock Chiron: Privacy-preserving machine learning as a service.
	\newblock {\em arXiv preprint arXiv:1803.05961}, 2018.
	
	\bibitem{Ryoan}
	Tyler Hunt, Zhiting Zhu, Yuanzhong Xu, Simon Peter, and Emmett Witchel.
	\newblock Ryoan: A distributed sandbox for untrusted computation on secret
	data.
	\newblock {\em ACM Trans. Comput. Syst.}, 35(4), December 2018.
	
	\bibitem{Gazelle}
	Chiraag Juvekar, Vinod Vaikuntanathan, and Anantha Chandrakasan.
	\newblock {GAZELLE: A Low Latency Framework for Secure Neural Network
		Inference}.
	\newblock In {\em Proceedings of the 27th USENIX Conference on Security
		Symposium}, SEC'18, pages 1651--1668, Berkeley, CA, USA, 2018. USENIX
	Association.
	
	\bibitem{kolesnikov2008improved}
	Vladimir Kolesnikov and Thomas Schneider.
	\newblock Improved garbled circuit: Free xor gates and applications.
	\newblock In {\em International Colloquium on Automata, Languages, and
		Programming}, pages 486--498. Springer, 2008.
	
	\bibitem{twn}
	Fengfu Li, Bo~Zhang, and Bin Liu.
	\newblock Ternary weight networks.
	\newblock {\em arXiv preprint arXiv:1605.04711}, 2016.
	
	\bibitem{DARTS}
	Hanxiao Liu, Karen Simonyan, and Yiming Yang.
	\newblock {DARTS}: Differentiable architecture search.
	\newblock In {\em International Conference on Learning Representations}, 2019.
	
	\bibitem{MiniONN}
	Jian Liu, Mika Juuti, Yao Lu, and N.~Asokan.
	\newblock Oblivious neural network predictions via {MiniONN} transformations.
	\newblock In {\em Proceedings of the 2017 ACM SIGSAC Conference on Computer and
		Communications Security}, CCS '17, pages 619--631, New York, NY, USA, 2017.
	ACM.
	
	\bibitem{Delphi}
	Pratyush Mishra, Ryan Lehmkuhl, Akshayaram Srinivasan, Wenting Zheng, and
	Raluca~Ada Popa.
	\newblock Delphi: A cryptographic inference service for neural networks.
	\newblock Cryptology ePrint Archive, Report 2020/050, 2020.
	\newblock \url{https://eprint.iacr.org/2020/050}.
	
	\bibitem{SecureML}
	P.~{Mohassel} and Y.~{Zhang}.
	\newblock {SecureML}: A system for scalable privacy-preserving machine
	learning.
	\newblock In {\em 2017 IEEE Symposium on Security and Privacy (SP)}, pages
	19--38, May 2017.
	
	\bibitem{ohrimenko2016oblivious}
	Olga Ohrimenko, Felix Schuster, C{\'e}dric Fournet, Aastha Mehta, Sebastian
	Nowozin, Kapil Vaswani, and Manuel Costa.
	\newblock Oblivious multi-party machine learning on trusted processors.
	\newblock In {\em USENIX Security Symposium}, 2016.
	
	\bibitem{paillier1999public}
	Pascal Paillier.
	\newblock Public-key cryptosystems based on composite degree residuosity
	classes.
	\newblock In Jacques Stern, editor, {\em Advances in Cryptology --- EUROCRYPT
		'99}, pages 223--238, Berlin, Heidelberg, 1999. Springer Berlin Heidelberg.
	
	\bibitem{enas}
	Hieu Pham, Melody Guan, Barret Zoph, Quoc Le, and Jeff Dean.
	\newblock Efficient neural architecture search via parameters sharing.
	\newblock In Jennifer Dy and Andreas Krause, editors, {\em Proceedings of the
		35th International Conference on Machine Learning}, volume~80 of {\em
		Proceedings of Machine Learning Research}, pages 4095--4104,
	Stockholmsmässan, Stockholm Sweden, 10--15 Jul 2018. PMLR.
	
	\bibitem{ObliviousTransfer_paper}
	Michael~O. Rabin.
	\newblock How to exchange secrets with oblivious transfer.
	\newblock Technical Report TR-81, Aiken Computation Lab, Harvard University,
	1981.
	
	\bibitem{XONN}
	M.~Sadegh Riazi, Mohammad Samragh, Hao Chen, Kim Laine, Kristin Lauter, and
	Farinaz Koushanfar.
	\newblock {XONN}: {XNOR}-based oblivious deep neural network inference.
	\newblock In {\em 28th {USENIX} Security Symposium ({USENIX} Security 19)},
	pages 1501--1518, Santa Clara, CA, August 2019. {USENIX} Association.
	
	\bibitem{Chameleon}
	M.~Sadegh Riazi, Christian Weinert, Oleksandr Tkachenko, Ebrahim~M. Songhori,
	Thomas Schneider, and Farinaz Koushanfar.
	\newblock {Chameleon: A Hybrid Secure Computation Framework for Machine
		Learning Applications}.
	\newblock In {\em Proceedings of the 2018 on Asia Conference on Computer and
		Communications Security}, ASIACCS '18, pages 707--721, New York, NY, USA,
	2018. ACM.
	
	\bibitem{DeepSecure}
	B.~D. {Rouhani}, M.~S. {Riazi}, and F.~{Koushanfar}.
	\newblock {DeepSecure}: Scalable provably-secure deep learning.
	\newblock In {\em 2018 55th {ACM/ESDA/IEEE} Design Automation Conference
		({DAC})}, June 2018.
	
	\bibitem{ShamirSS}
	Adi Shamir.
	\newblock How to share a secret.
	\newblock {\em Commun. ACM}, 22(11):612--613, Nov 1979.
	
	\bibitem{shokri2017membership}
	Reza Shokri, Marco Stronati, Congzheng Song, and Vitaly Shmatikov.
	\newblock Membership inference attacks against machine learning models.
	\newblock In {\em Security and Privacy (SP), 2017 IEEE Symposium on}, 2017.
	
	\bibitem{TinyGarble_paper}
	E.~M. {Songhori}, S.~U. {Hussain}, A.~{Sadeghi}, T.~{Schneider}, and
	F.~{Koushanfar}.
	\newblock {TinyGarble: Highly Compressed and Scalable Sequential Garbled
		Circuits}.
	\newblock In {\em 2015 IEEE Symposium on Security and Privacy}, pages 411--428,
	May 2015.
	
	\bibitem{tramer2016stealing}
	Florian Tram{\`e}r, Fan Zhang, Ari Juels, Michael~K Reiter, and Thomas
	Ristenpart.
	\newblock Stealing machine learning models via prediction {APIs}.
	\newblock In {\em USENIX Security}, 2016.
	
	\bibitem{Slalom}
	Florian Tramèr and Dan Boneh.
	\newblock Slalom: Fast, verifiable and private execution of neural networks in
	trusted hardware, 2018.
	
	\bibitem{van2010fully}
	Marten van Dijk, Craig Gentry, Shai Halevi, and Vinod Vaikuntanathan.
	\newblock Fully homomorphic encryption over the integers.
	\newblock In Henri Gilbert, editor, {\em Advances in Cryptology -- EUROCRYPT
		2010}, pages 24--43, Berlin, Heidelberg, 2010. Springer Berlin Heidelberg.
	
	\bibitem{Crypto-nets}
	Pengtao Xie, Misha Bilenko, Tom Finley, Ran Gilad-Bachrach, Kristin~E. Lauter,
	and Michael Naehrig.
	\newblock {Crypto-Nets}: Neural networks over encrypted data.
	\newblock {\em CoRR}, abs/1412.6181, 2014.
	
	\bibitem{yao1982protocols}
	Andrew~C. Yao.
	\newblock Protocols for secure computations.
	\newblock In {\em Proceedings of the 23rd Annual Symposium on Foundations of
		Computer Science}, SFCS ’82, page 160–164, USA, 1982. IEEE Computer
	Society.
	
	\bibitem{YaoGC}
	Andrew~C. {Yao}.
	\newblock How to generate and exchange secrets.
	\newblock In {\em 27th Annual Symposium on Foundations of Computer Science
		(SCFS 1986)}, pages 162--167, Oct 1986.
	
	\bibitem{nas_rl}
	Barret Zoph and Quoc~V. Le.
	\newblock Neural architecture search with reinforcement learning.
	\newblock {\em CoRR}, abs/1611.01578, 2016.
	
\end{thebibliography}

\onecolumn

\newpage
\appendix
\section{Specifications of the Model Architectures} 
\label{Appendix:architectures}

\begin{table*}[h]
    \centering
    \caption{Model architectures used in our experiments. Models m1-6 are used in the prior work on MNIST and CIFAR10 datasets, and we constructed the \method{} models using our regularized architecture search algorithm.}
    \label{tab:architectures}
    \footnotesize
    \setlength\tabcolsep{3 pt}
\begin{tabular}{lllllllllllllll} 
    \cline{1-3}\cline{5-7}\cline{9-11}\cline{13-15}
    \textbf{}  & \textbf{Type} & \begin{tabular}[c]{@{}l@{}}\textbf{Kernels/}\\\textbf{ Nodes} \end{tabular} &  &    & \textbf{Type} & \begin{tabular}[c]{@{}l@{}}\textbf{Kernels/}\\\textbf{Nodes} \end{tabular} &  &    & \textbf{Type}    & \begin{tabular}[c]{@{}l@{}}\textbf{Kernels/}\\\textbf{Nodes}\end{tabular} &  & & \textbf{Type} & \begin{tabular}[c]{@{}l@{}}\textbf{Kernels/}\\\textbf{Nodes} \end{tabular}    \\ 
    \cline{1-3}\cline{5-7}\cline{9-11}\cline{13-15}
    & &  &  &    & &  &  &    &  &  &  & & & \\
    \multicolumn{3}{c}{\textbf{MNIST (m1)}} &  & \multicolumn{3}{c}{\textbf{CIFAR10 (m5)} }  &  & \multicolumn{3}{c}{\textbf{CIFAR10 (\method)} }  &  & \multicolumn{3}{c}{\textbf{CIFAR10 (\method)} } \\ 
    \cline{1-3}\cline{5-7}\cline{9-11}
    1  & FC  & 128 &  & 1  & CONV $3\times3$ & 16 &  & \multicolumn{3}{c}{\multirow{2}{*}{\begin{tabular}[c]{@{}c@{}}$\lambda=0$, No. of cells = 3,\\No. of operations per cell = 4 \end{tabular}}} &  & \multicolumn{3}{c}{\multirow{2}{*}{\begin{tabular}[c]{@{}c@{}}$\lambda=0.6$, No. of cells = 3,\\ No. of operations per cell = 4 \end{tabular}}}  \\
    2  & FC  & 128 &  & 2  & CONV $3\times3$ & 16 &  & \multicolumn{3}{c}{} &  & \multicolumn{3}{c}{} \\ 
    \cline{13-15}
    3  & FC  & 10 &  & 3  & CONV $3\times3$ & 16 &  & 1  & CONV $5\times5$ & 16    &  & 1    & CONV $3\times3$ & 16  \\ 
    \cline{1-3}
    & &  &  & 4  & MAXPOOL $2\times2$  & -  &  & 2  & MAXPOOL~$2\times2$~   & - &  & 2    & CONV $3\times3$ & 16  \\
    \multicolumn{3}{c}{\textbf{MNIST (m2)} } &  & 5  & CONV $3\times3$ & 32 &  & 3  & CONV $5\times5$ & 16    &  & 3    & CONV $3\times3$ & 16  \\ 
    \cline{1-3}
    1  & CONV $5\times5$ & 5  &  & 6  & CONV $3\times3$ & 32 &  & 4  & CONV $5\times5$ & 16    &  & 4    & MAXPOOL $2\times2$   & -   \\
    2  & FC  & 100 &  & 7  & CONV $3\times3$ & 32 &  & 5  & CONV $5\times5$ & 32    &  & 5    & CONV $3\times3$ & 32  \\
    3  & FC  & 10 &  & 8  & MAXPOOL $2\times2$  & -  &  & 6  & MAXPOOL~$2\times2$~   & - &  & 6    & CONV $3\times3$ & 32  \\ 
    \cline{1-3}
    & &  &  & 9  & CONV $3\times3$ & 48 &  & 7  & CONV $5\times5$ & 32    &  & 7    & CONV $3\times3$ & 32  \\
    \multicolumn{3}{c}{\textbf{MNIST (m3)} } &  & 10 & CONV $3\times3$ & 48 &  & 8  & CONV $5\times5$ & 32    &  & 8    & MAXPOOL $2\times2$   & -   \\ 
    \cline{1-3}
    1  & CONV $5\times5$ & 16 &  & 11 & CONV $3\times3$ & 64 &  & 9  & CONV $5\times5$ & 64    &  & 9    & CONV $3\times3$ & 64  \\
    2  & MAXPOOL $2\times2$  & -  &  & 12 & MAXPOOL $2\times2$  & -  &  & 10 & MAXPOOL~$2\times2$~   & - &  & 10   & CONV $3\times3$ & 64  \\
    3  & CONV $5\times5$ & 16 &  & 13 & FC  & 10 &  & 11 & CONV $5\times5$ & 64    &  & 11   & CONV $3\times3$ & 64  \\ 
    \cline{5-7}
    4  & MAXPOOL $2\times2$  & -  &  &    & &  &  & 12 & CONV $5\times5$ & 64    &  & 12   & MAXPOOL $2\times2$   & -   \\
    5  & FC  & 100 &  & \multicolumn{3}{c}{\textbf{CIFAR10 (m6)} }  &  & 13 & FC    & 10    &  & 13   & FC   & 10  \\ 
    \cline{5-7}\cline{9-11}\cline{13-15}
    6  & FC  & 10 &  & 1  & CONV $3\times3$ & 16 &  &    &  &  &  & & & \\ 
    \cline{1-3}
    & &  &  & 2  & CONV $3\times3$ & 32 &  & \multicolumn{3}{c}{\textbf{MNIST (\method)} }   &  & \multicolumn{3}{c}{\textbf{MNIST (\method)} } \\
    \multicolumn{3}{c}{\textbf{CIFAR10 (m4)} } &  & 3  & CONV $3\times3$ & 32 &  & \multicolumn{3}{c}{\multirow{2}{*}{\begin{tabular}[c]{@{}c@{}}$\lambda=0$, No. of cells = 1,\\No. of operations per cell = 4 \end{tabular}}} &  & \multicolumn{3}{c}{\multirow{2}{*}{\begin{tabular}[c]{@{}c@{}}$\lambda=0.6$, No. of cells = 1,\\ No. of operations per cell = 4 \end{tabular}}}  \\ 
    \cline{1-3}
    1  & CONV $3\times3$ & 64 &  & 4  & MAXPOOL $2\times2$  & -  &  & \multicolumn{3}{c}{} &  & \multicolumn{3}{c}{} \\ 
    \cline{9-11}\cline{13-15}
    2  & CONV $3\times3$ & 64 &  & 5  & CONV $3\times3$ & 48 &  & 1  & CONV~$5\times5$  & 16    &  & 1    & CONV $3\times3$ & 16  \\
    3  & MAXPOOL $2\times2$  & -  &  & 6  & CONV $3\times3$ & 64 &  & 2  & CONV~$5\times5$  & 16    &  & 2    & CONV $3\times3$ & 16  \\
    4  & CONV $3\times3$ & 64 &  & 7  & CONV $3\times3$ & 80 &  & 3  & CONV~$5\times5$  & 16    &  & 3    & MAXPOOL $2\times2$   & -   \\
    5  & CONV $3\times3$ & 64 &  & 8  & MAXPOOL $2\times2$  & -  &  & 4  & CONV~$5\times5$  & 16    &  & 4    & CONV $5\times5$ & 16  \\
    6  & MAXPOOL $2\times2$  & -  &  & 9  & CONV $3\times3$ & 96 &  & 5  & FC    & 100   &  & 5    & FC   & 100 \\
    7  & CONV $3\times3$ & 64 &  & 10 & CONV $3\times3$ & 96 &  & 6  & FC    & 10    &  & 6    & FC   & 10  \\ 
    \cline{9-11}\cline{13-15}
    8  & CONV $1\times1$ & 64 &  & 11 & CONV $3\times3$ & 128 &  &    &  &  &  & & & \\
    9  & CONV $1\times1$ & 16 &  & 12 & MAXPOOL $2\times2$  & -  &  &    &  &  &  & \multicolumn{1}{c}{} & \multicolumn{1}{c}{} & \multicolumn{1}{c}{}  \\
    10 & FC  & 10 &  & 13 & FC  & 10 &  &    &  &  &  & & & \\
    \cline{1-3}\cline{5-7}
\end{tabular}
\end{table*}

\end{document}